\documentclass[prl,twocolumn,showpacs,lengthcheck]{revtex4}

\usepackage{amsmath}
\usepackage{amssymb}
\usepackage{amsthm}
\usepackage{graphicx}

\newcommand{\uinvnorm}{|\kern-2pt|\kern-2pt|}

\theoremstyle{plain}

\theoremstyle{definition}

\theoremstyle{remark}

\begin{document}

\bibliographystyle{apsrev}

\title{The Dynamics of 1D Quantum Spin Systems Can Be Approximated Efficiently}

\author{Tobias J.\ Osborne}
\email[]{Tobias.Osborne@rhul.ac.uk} \affiliation{Department of
Mathematics, Royal Holloway University of London, Egham, Surrey TW20
0EX, UK}

\date{\today}

\begin{abstract}
In this Letter we show that an arbitrarily good approximation to the
propagator $e^{itH}$ for a $1$D lattice of $n$ quantum spins with
hamiltonian $H$ may be obtained with polynomial computational
resources in $n$ and the error $\epsilon$, and exponential resources
in $|t|$. Our proof makes use of the finitely correlated
state/matrix product state formalism exploited by numerical
renormalisation group algorithms like the density matrix
renormalisation group. There are two immediate consequences of this
result. The first is that Vidal's time-dependent density matrix
renormalisation group will require only polynomial resources to
simulate $1$D quantum spin systems for logarithmic $|t|$. The second
consequence is that continuous-time $1$D quantum circuits with
logarithmic $|t|$ can be simulated efficiently on a classical
computer, despite the fact that, after discretisation, such circuits
are of polynomial depth.
\end{abstract}


\maketitle

The kinematics and dynamics of quantum lattice systems are
strongly constrained by a key physical requirement, namely, the
\emph{locality of interactions}. For example, consider a
collection of $n$ distinguishable spin-$\frac12$ systems which
interact according to nearest-neighbour interactions: a counting
argument quickly reveals that such local hamiltonians occupy an
extremely small fraction of the space of general hamiltonians.
Thus, intuitively, we would expect this nongeneric constraint
would manifest itself strongly in the structure of the eigenvalues
and eigenstates for local hamiltonians. This is indeed the case,
but it is still far from obvious exactly how to best quantify this
constraint.

A number of methods to systematically quantify the eigenstates and
matrix functions of local hamiltonians have been developed. Perhaps
the most successful scheme in recent years has been the technology
of \emph{finitely correlated quantum states} (FCS)
\cite{fannes:1992a, richter:1996}. (Finitely correlated states are
also known as \emph{matrix product states} (MPS) in one dimension
and \emph{tensor product states} or \emph{projected entangled-pair
states} in two and higher dimensions \cite{verstraete:2004a}. The
key feature of a finitely correlated state is that, as the name
suggests, separated regions are weakly correlated. In addition, any
state which does not exhibit too much correlation between separated
subsystems can be well approximated by a finitely correlated state
\cite{verstraete:2005a}, \cite{endnote22}.

Finitely correlated states are nothing more than a convenient
representation for vectors in tensor-product hilbert spaces.
However, the utility of this particular representation is that for
those states with bounded or limited correlations it is often
extremely efficient (in $n$) to extract local properties, such as
expectation values of local operators.

The utility of the FCS representation as a means to calculate
local properties of $1$D quantum lattice systems has been
spectacularly demonstrated by the development of the \emph{density
matrix renormalisation group} (DMRG). (See
\cite{schollwoeck:2005a} and references therein for a description
of the DMRG and related algorithms.) The DMRG provides an
apparently efficient computational recipe to obtain an
approximation to the ground state and low-energy eigenstates for
$1$D quantum lattice systems as FCS vectors. The DMRG is an
extremely flexible method and has been recently extended to apply
to a diverse number of situations, such as the calculation of
short-time dynamics \cite{vidal:2003a, vidal:2003b}, dissipation
\cite{verstraete:2004b, zwolak:2004a}, disordered systems
\cite{paredes:2005a}, eigenstates with definite momentum
\cite{porras:2005a}, and, recently, higher dimensions
\cite{verstraete:2004a}.

Perhaps one of the most exciting recent results in the study of
the DMRG has been the development of an algorithm to simulate the
real-time dynamics of $1$D quantum spin systems \cite{vidal:2003a,
vidal:2003b, white:2004a}. The efficiency of this algorithm is
predicated on the condition that the dynamics of the spin system
do not create too much long-range quantum entanglement
\cite{vidal:2003a, vidal:2003b}. While it appears that, in
practice, this condition is always fulfilled for small times, it
is currently unclear if it is true for \emph{all} $1$D local
quantum spin systems.

There are at least two reasons why it is interesting to study the
theoretical worst-case computational costs of the time-dependent
DMRG. The first is that the computational complexity of the DMRG and
related algorithms is currently unknown except for when applied to a
handful of singular integrable models \cite{peschel:1999a}. An
assessment of the theoretical worst-case computational complexity of
the DMRG in any other circumstance would allow one to certify
\emph{a priori} the accuracy of the DMRG versus computational cost.
The second reason is related to computational power of quantum
computers (see \cite{nielsen:2000a} for a detailed description of
quantum computation and a number of quantum algorithms including
quantum simulation algorithms). The time-dependent DMRG provides a
way to simulate quantum computers running quantum algorithms. A
careful theoretical worst-case complexity analysis would potentially
give us an insight into what quantum computations can and can't be
simulated efficiently on a classical computer.

The most naive way to study the computational complexity of the
time-dependent DMRG is to directly study the storage costs of
representing the propagator $e^{itH}$ when it is approximated by the
Lie-Trotter expansion $e^{itH} \approx
(e^{i\frac{t}{m}A}e^{i\frac{t}{m}B})^m$, for some large $m$
\cite{endnote23}. Unfortunately, however, a careful analysis of the
error scaling with $m$ for this representation shows that the
worst-case storage cost might be exponential, even for $|t|$ which
scales as a constant with $n$. This, in turn, implies that quantum
circuits simulating $1$D dynamics using the Lie-Trotter expansion
have a depth that scales at least linearly with $n$. If we want to
simulate such methods efficiently we need a more sophisticated
technique to obtain a representation for the propagator.

\begin{figure}
\center
\includegraphics{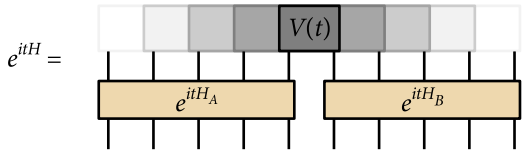}
\caption{Decomposing the propagator: for a given time $t$ the
propagator $e^{itH}$ may be written as a product of $e^{it(H_A +
H_B)}$ and $V(t)$, where $H_A$ (respectively, $H_B$) is the
hamiltonian with interaction terms only between spins in region $A$
(respectively, between spins in region $B$) and $V(t)$ is a unitary
operator which ``patches up" the error due to approximating
$e^{itH}$ with $e^{it(H_A + H_B)}$. Because of the UV cutoff given
by the lattice, information propagation in quantum spin systems is
limited by an effective ``speed of light". Hence the unitary $V(t)$
interacts spins in $A$ with spins in $B$ successively weaker further
from the boundary between $A$ and $B$.}\label{fig:uvdec}
\end{figure}

\begin{figure}
\center
\includegraphics{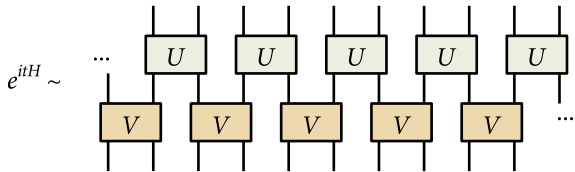}
\caption{The approximate quantum cellular automata decomposition for
$e^{itH}$ for time $t$. Each of the unitary operators $U$ and $V$
acts on at most $c_0 |t|$ spins, where $c_0$ is a
constant.}\label{fig:qca}
\end{figure}

In this Letter we show that the propagator $e^{itH}$ for an
\emph{arbitrary} $1$D quantum system is well-approximated by a
finitely correlated state vector in the Hilbert space of operators
using resources that scale polynomially with $n$ and exponentially
with $|t|$. Because our proof is constructive we obtain an efficient
algorithm, closely related to the DMRG, to obtain this
representation.

The argument we describe in this Letter can be understood by
appealing to the following physical intuition. The dynamics of any
1D quantum spin system are constrained by the ultraviolet cutoff
induced by the lattice spacing between the spins. This cutoff
induces a bound on the speed at which information can propagate in
such systems, an intuition which is precisely quantified by the
\emph{Lieb-Robinson bound} \cite{lieb:1972a}. We exploit this bound
on information propagation to provide two different decompositions
of the propagator $e^{itH}$. In the first decomposition we partition
the chain into two contiguous pieces $A$ and $B$ and approximate the
dynamics $e^{itH}$ by $e^{it(H_A+H_B)}$, where $H_A$ is the
hamiltonian which includes interaction terms only between spins in
$A$ (and similarly for $H_B$). Obviously this approximation is not
exact: at the cut point between $A$ and $B$ there will be
substantial errors. However, as a consequence of the bound on
information propagation, the difference between the way $e^{itH}$
and $e^{it(H_A+H_B)}$ act on spins far from the cut will become
small (information about the cut doesn't have time to propagate too
far away). We can patch up the difference between the two
propagators by introducing a new unitary operator $V(t)$ which acts
across the boundary: $e^{itH} = e^{it(H_A+H_B)}V(t)$. The
Lieb-Robinson bound then tells us that $V(t)$ acts progressively
weaker on spins far from the cut (see Fig.~\ref{fig:uvdec} for
discussion and illustration of this). This, in turn, allows us to
approximate $V(t)$ with a unitary $V'(t)$ which acts nontrivially
only on a \emph{finite} number of spins $\Omega$ around the
boundary. If $|\Omega|$ is bigger than $|t|$ then this approximation
improves exponentially fast in $|\Omega|$.

We obtain the second decomposition for $e^{itH}$ by first fixing $t$
and then moving along the chain $\mathcal{C}$ and introducing a cut
after approximately $|t|$ spins (which we call $\Lambda_1$) and then
patching it up with a unitary $V_1(t)$ which acts nontrivially only
on a finite number of spins across the boundary (this is the unitary
constructed in the previous paragraph). Thus we have $e^{itH}\sim
V_1(t) e^{itH_{\Lambda_1}}\otimes e^{itH_{\mathcal{C}\setminus
\Lambda_1}}$. Then we recursively apply this procedure to
$e^{itH_{\mathcal{C}\setminus \Lambda_1}}$ until we arrive at the
approximate decomposition illustrated in Fig.~\ref{fig:qca}.

We will, for the sake of clarity, describe our results mainly for a
finite chain $\mathcal{C}$ of $n$ distinguishable spin-$1/2$
particles. The family $H$ of local hamiltonians we focus on (which
implicitly depends on $n$) is defined by $H = \sum_{j=0}^{n-2} h_j$,
where $h_j$ acts nontrivially only on spins $j$ and $j+1$. We set
the energy scale by assuming that $\|h_j\|$ scales as a constant
with $n$ for all $j = 0,1, \ldots, n-1$, where $\|\cdot\|$ denotes
the operator norm. The interaction terms $h_j$ may depend on time:
$h_j = h_j(t)$. We can easily accommodate next-nearest neighbour
interactions etc.\ by blocking sites and thinking of the blocks as
new (larger) spins. However this can only be done a constant number
of times: the quality of our approximation will decrease
exponentially with the number of such blockings. We do not assume
translational invariance.

The crucial idea underlying our approach is that a good
approximation to the propagator $e^{itH}$ for a local $1$D quantum
spin lattice system can be obtained and stored efficiently (i.e.\
with polynomial resources in $n$) with a classical computer for $|t|
\le c\log(n)$, where $c$ is some constant. The way we do this is to
use a specific representation for the approximation, namely as a
\emph{finitely correlated state vector}. What we mean by this is
that we represent an operator $W$ in the following fashion
\begin{equation}\label{eq:fcsdef}
W = \sum_{\boldsymbol{\alpha}\in
Q_n}\mathbf{A}^{\alpha_0}\mathbf{A}^{\alpha_1}\cdots
\mathbf{A}^{\alpha_{n-1}} \sigma^{\alpha_0}\otimes
\sigma^{\alpha_1} \otimes \cdots \otimes \sigma^{\alpha_{n-1}},
\end{equation}
where $Q_n = \{0,1,2,3\}^{\times n}$, $\mathbf{A}^{\alpha_0}$
(respectively, $\mathbf{A}^{\alpha_{n-1}}$) are a collection of four
row vectors of size $D_0$ (respectively, four column vectors of size
$C_{n-1}$), $\mathbf{A}^{\alpha_j}$ are four $C_j\times D_j$ sized
matrices, for $1 \le j < n-1$, and $\sigma^{\alpha}$ is the vector
of Pauli operators. Note that $C_{j+1} = D_{j}$. The dimensions
$C_j$ and $D_j$ are called the auxiliary dimensions for site $j$. It
is clear that if the sizes of the auxiliary dimensions are bounded
by polynomials in $n$, i.e.\ if $C_j \le \text{poly}(n)$ and $D_j
\le \text{poly}(n)$, then the operator $W$ can be stored with
polynomial resources in $n$. Also note that all operators can be
represented exactly as in Eq.~(\ref{eq:fcsdef}) by taking the
auxiliary dimensions to be large enough: $C_j = D_j = 2^n$ suffices
\cite{fannes:1992a, verstraete:2004c}.

We begin by showing how to obtain an approximate decomposition of
the propagator $e^{itH}$ as a product
\begin{equation}\label{eq:uvmtimedec}
e^{itH} = \left(\bigotimes_{j = 0}^{n/|\Omega|-1}
U_{\Omega_j}(t)\right)\left(\bigotimes_{j = 0}^{n/|\Omega|}
V_{\Omega_j'}(t)\right) + O(\epsilon),
\end{equation}
where we have two partitions $\mathcal{P}_1$ and $\mathcal{P}_2$ of
the chain $\mathcal{C}$ into $n/|\Omega|$ (respectively
$n/|\Omega|+1$) contiguous blocks of $\le|\Omega|$ spins. The first
set is denoted $\mathcal{P}_1 = \{\Omega_j\}$. The second set
$\mathcal{P}_2 = \{\Omega_j'\}$ is the set of blocks which are just
translates of those in $\mathcal{P}_1$ by $|\Omega|/2$ sites
\cite{endnote24}. The operators $U_{\Omega_j}$ (respectively,
$V_{\Omega_j'}$) are unitary operators which act nontrivially only
on $\Omega_j$ (respectively, ${\Omega_j'}$). We call such a
decomposition an \emph{approximate quantum cellular automata
decomposition} (or, simply, an $\epsilon$-QCA decomposition) because
$e^{itH}$ is exactly a Margolus-partitioned QCA update rule (see
\cite{schumacher:2004a} for a description of QCA's). Then we show
that an $\epsilon$-QCA decomposition implies that $e^{itH}$ is
well-approximated by a FCS vector with auxiliary dimension
$2^{|\Omega|}$. To reduce the error to $\epsilon$ we require
$|\Omega| \ge c_0 |t| + c_1\log(n/\epsilon)$, where $c_0$ and $c_1$
are constants. This decomposition is illustrated in
Fig.~\ref{fig:qca}.

Consider the unitary operator
\begin{equation}\label{eq:basicdec}
V(t) = (e^{-itH_{\mathcal{C}\setminus\Lambda}}\otimes
e^{-itH_{\Lambda}})e^{itH}.
\end{equation}
As we described in the introduction, for small $|t|$, and for sites
far enough away from the boundary $\partial \Lambda$ between
$\mathcal{C}\setminus\Lambda$ and $\Lambda$, this operator ought to
be close to the identity. Therefore we argue that, as an operator,
$V(t)$ ought to be expressible as $V(t) \approx
\mathbb{I}_{\mathcal{C}\setminus\Omega} \otimes V'_\Omega(t).$

To quantify this statement we study the differential equation that
$V(t)$ satisfies:
\begin{equation}\label{eq:unischro1}
\frac{dV(t)}{dt} = ie^{-it(H-h_I)}h_Ie^{itH} = i V(t)
\tau_t^{H}(h_I),
\end{equation}
where $h_I$ is the interaction term that bridges the left- and
right-hand side of the chain, and $\tau_{t}^{B}(A) =
e^{-itB}Ae^{itB}$. Thus we see that $V(t)$ is generated by
time-dependent unitary dynamics due to the effective hamiltonian
$L(t) = \tau_t^{H}(h_I)$, and we write
\begin{equation}
V(t) = \mathcal{T} e^{i\int^t_0 L(s)\,ds},
\end{equation}
where $\mathcal{T}$ denotes time-ordering.

The time-dependent effective hamiltonian $L(t)$, and hence $V(t)$,
acts nontrivially on all of the sites in the chain. We now show
that, for small $|t|$, $L(t)$ may be well-approximated by an
operator which acts nontrivially on only a handful of sites near the
boundary $\partial\Lambda$ between the left- and right-hand sides of
the chain. To do this, we construct the following approximation to
$L(t)$:
\begin{equation}
L'(t) = \tau_{t}^{H_\Omega}(h_I),
\end{equation}
where $\Omega$ is a contiguous block of sites centred on $\partial
\Lambda$, and $H_\Omega$ contains only those interaction terms $h_j$
in $H$ which interact only spins contained in $\Omega$. We  now show
that, for small enough $|t|$ and large enough $\Omega$ containing
$H_I$, $\|L(t)-L'(t)\| < \epsilon$, for some prespecified
$\epsilon$.

To show that $L'(t)$ is a good approximation to $L(t)$ we must
establish that $\|\tau_t^{H}(h_I) - \tau_t^{H_\Omega}(h_I)\|$ is
small. A bound on such a quantity is known as a \emph{Lieb-Robinson
bound} \cite{lieb:1972a, hastings:2004a, nachtergaele:2005a,
hastings:2005b} (see \cite{osborne:2006b} for a simple direct
proof). The strongest (and easiest to prove) such bound reads
\begin{equation*}
\| \tau_t^{H}(h_I)-\tau_t^{H_{\Omega}}(h_I) \| \le
\sum_{l=|\Omega|}^\infty \delta_l|t|^l/l! \le \omega
e^{\kappa|t|}e^{- \mu|\Omega|},
\end{equation*}
where $\delta_l =  \|h_I\| 2^l\|h\|^l$, $\|h\| = \max_j \|h_j\|$,
and $\kappa$, $\mu$, and $\omega$ are constants. Thus we find
\begin{equation}\label{eq:lineq}
\|L(t)-L'(t)\| = {\omega}e^{\kappa|t|}e^{- \mu|\Omega|}.
\end{equation}
In this way we see that we can reduce the operator norm difference
between $L(t)$ and $L'(t)$ exponentially fast in the size
$|\Omega|$ of the region $\Omega$.

We now define a new unitary operator $V'(t)$ --- which is meant to
approximate $V(t)$ --- as the unitary operator generated by the
time-dependent hamiltonian $L'(t)$:
\begin{equation}\label{eq:unischro2}
\frac{dV'(t)}{dt} =  iV'(t)L'(t).
\end{equation}
Because $L'(t)$ acts nontrivially only on $\Omega$, $V'(t)$ is a
unitary operator which acts nontrivially only on $\Omega$ and it
acts as an identity elsewhere. In order to see how accurate $V(t)$
is as an approximation to $V'(t)$ we now bound the error
$\|V(t)-V'(t)\|$.

To show that $V(t)$ and $V'(t)$ are close for some time period we
integrate the differential equations (\ref{eq:unischro1}) and
(\ref{eq:unischro2}). We do this by making use of the Lie-Trotter
expansion
\begin{align}
V(t) &= \lim_{m\rightarrow \infty} \prod_{j=0}^{m-1} e^{iL(\frac{jt}{m})\frac{t}{m}} \\
V'(t) &= \lim_{m\rightarrow \infty} \prod_{j=0}^{m-1}
e^{iL'(\frac{jt}{m})\frac{t}{m}},
\end{align}
applying the triangle inequality several times, and taking the
limit ${m\rightarrow\infty}$. This gives us the fundamental
estimate
\begin{equation}
\|V(t)-V'(t)\| \le \int_{0}^{|t|} \|L(s)-L'(s)\| ds.
\end{equation}
Substituting (\ref{eq:lineq}) and redefining constants gives us
\begin{equation}\label{eq:basicest}
\|V(t)-V'(t)\| \le \omega e^{\kappa |t|}e^{-\mu |\Omega|},
\end{equation}
where $\omega$, $\kappa$, and $\mu$ are constants independent of
$n$.

Our final result is now the following. Rearranging
(\ref{eq:basicdec}) and using the estimate (\ref{eq:basicest})
gives us
\begin{equation}
e^{itH} = (e^{itH_{\mathcal{C}\setminus\Lambda}}\otimes
e^{itH_{\Lambda}})V'(t) + \epsilon,
\end{equation}
where $V'(t)$ acts nontrivially only on a contiguous block
$\Omega$ of spins of size $|\Omega|$. Iterating this procedure by
cutting $\Lambda$ into two pieces etc.\ give us the final
$\epsilon$-QCA decomposition
\begin{equation}
e^{itH} = \left(\bigotimes_{j = 0}^{n/|\Omega|-1}
U_{\Omega_j}(t)\right)\left(\bigotimes_{j = 0}^{n/|\Omega|}
V_{\Omega_j'}(t)\right) + O(\epsilon),
\end{equation}
where $|\Omega| = O(c_0|t|+c_1\log(n/\epsilon))$, for some
constants $c_0$ and $c_1$.

It is now relatively straightforward to show that an
$\epsilon$-QCA decomposition gives rise to an efficient FCS vector
representation once we recognise that the expression
\begin{equation}
\mathcal{U}(t) = \left(\bigotimes_{j = 0}^{n/|\Omega|-1}
U_{\Omega_j}(t)\right)
\end{equation}
is a FCS vector with auxiliary dimension $2^{|\Omega|}$ The way to
see this is to note that, in the standard operator basis,
\begin{equation}
\begin{split}
U(t) &= \prod_{j = 0}^{n/|\Omega|-1} \left(\sum_{\mathbf{k}_j\in
Q_{|\Omega|}} c_{\mathbf{k}_j}(t)
\sigma_{\Omega_j}^{\mathbf{k}_j}\right) \\
&= \sum_{\mathbf{k}}c_{\mathbf{k}_0}(t)c_{\mathbf{k}_1}(t)\cdots
c_{\mathbf{k}_{n/|\Omega|-1}}(t) \sigma^{\mathbf{k}},
\end{split}
\end{equation}
which is a FCS vector representation with maximum auxiliary
dimension $2^{|\Omega|}$.

Given that
\begin{equation}
\mathcal{V}(t) = \left(\bigotimes_{j = 0}^{n/|\Omega|}
V_{\Omega_j'}(t)\right)
\end{equation}
is also expressible exactly as a FCS with maximum auxiliary
dimension $2^{|\Omega|}$ and using the result that the product of
two FCS vector operators with maximum auxiliary dimensions $D_1$
and $D_2$ admits a FCS vector expression with maximum auxiliary
dimensions $D_1D_2$ we obtain the final result that our
approximation $V'(t) = \mathcal{U}(t)\mathcal{V}(t)$ to $V(t)$ is
expressible exactly as a FCS with maximum auxiliary dimension $\le
2^{2|\Omega|}$. If $|t|$ scales as a constant, or logarithmically,
with $n$ then we learn that the FCS representation requires only
polynomial resources in $n$.

We have shown how the propagator $e^{itH}$ for a $1$D system of
quantum spins can be efficiently obtained and represented using a
classical computer. There are many consequences of this
representation. The first and most obvious is that the
time-dependent DMRG will, in the theoretical worst case, require
only polynomial computational resources to simulate time evolution
for constant times.

\begin{acknowledgements}
I would especially like to thank to thank Ignacio Cirac, Toby
Cubitt, Jens Eisert, Henry Haselgrove, Nick Jones, Julia Kempe,
Llu{\'\i}s Masanes, David P\'erez-Garc{\'\i}a, Jiannis Pachos,
Diego Porras, Tony Short, Frank Verstraete, Guifr{\'e} Vidal,
Andreas Winter, and Michael Wolf for their numerous helpful
comments and suggestions. I am grateful to the EU for support for
this research under the IST project RESQ and also to the UK EPSRC
through the grant QIPIRC.
\end{acknowledgements}


\begin{thebibliography}{21}
\bibitem{endnote22}{It is worth noting that any quantum state can be
written as a FCS, however, the more correlated the state, the more
expensive (in terms of memory usage) this becomes.}
\bibitem{endnote23}{The Lie-Trotter expansion is the key ingredient
exploited to simulate dynamics in the time-dependent DMRG.}
\bibitem{endnote24}{At each end of the chain the second partition
$\mathcal {P}_2$ has a block of size $|\Omega |/2$}
\end{thebibliography}
\end{document}